\documentclass[%
 reprint,
 bibnotes,
 amsmath,amssymb,
 aps,
 prd
]{revtex4-2}

\usepackage{graphicx,color}
 \usepackage{bm}% bold math
   \usepackage{amsmath}
    \usepackage{amssymb}    
     \usepackage{pifont}
\usepackage{tikz}
\usetikzlibrary{decorations.pathmorphing}
\usetikzlibrary{decorations.markings}
\usetikzlibrary{calc}
\usetikzlibrary{math}
\usetikzlibrary{patterns}
\usepackage{braket}
\usepackage{standalone}
\usepackage{slashed}
\usepackage{multirow}
\usepackage{hhline}
\usepackage{colortbl}

\newcommand{\nn}{\nonumber}

\newcommand{\ot}{\leftarrow}

\renewcommand{\(}{\left(}
\renewcommand{\)}{\right)}
\renewcommand{\[}{\left[}
\renewcommand{\]}{\right]}

\begin{document} 

\title{Study of worm-gear-T function $g_{1T}$ with semi-inclusive DIS data}
\author{Malin Horstmann}
\author{Andreas Sch\"afer}
\affiliation{Institut f\"ur Theoretische Physik, Universit\"at Regensburg, D-93040 Regensburg, Germany}

\author{Alexey Vladimirov}
\affiliation{Dpto. de F\'isica Te\'orica \& IPARCOS, Universidad Complutense de Madrid, E-28040 Madrid, Spain}

\begin{abstract}
The worm-gear-T function parameterizes the probability to find a longitudinally polarized quark inside a transversely polarized hadron. We extract it from data on polarized semi-inclusive deep-inelastic scattering (SIDIS) measured at COMPASS and HERMES. As theoretical model we use a Wandzura-Wilczek-type approximation at next-to-leading order. We find that at present, the data quality is insufficient to determine worm-gear-T function faithfully. We also provide predictions for the transverse single-spin asymmetry associated with the worm-gear-T function, which could be measured in weak-boson production at STAR.
\end{abstract}

\maketitle 

\section{Introduction}

The complexity of the internal hadron's structure cannot be described entirely in terms of one-dimensional collinear distributions but requires a multi-dimensional formalism.  The transverse-momentum dependent (TMD) factorization formalism \cite{Mulders:1995dh, Collins:2011zzd} is targeted at the understanding of this complex structure in terms of the partons' three-dimensional momentum distributions (TMD distributions). During the last years, the importance of internal transverse momentum has been confirmed in multiple cases, see examples in refs.\cite{Bacchetta:2017gcc, Scimemi:2017etj, Scimemi:2019cmh, Vladimirov:2019bfa, Bury:2020vhj, Echevarria:2020hpy, Bastami:2020asv, Cammarota:2020qcw, Bhattacharya:2021twu, Bacchetta:2022awv}. There are eight leading twist TMD distributions \cite{Bacchetta:2006tn} and only a few of them are known to some degree of certainty. In this contribution, we study the worm-gear-T function (wgt-function), denoted by $g_{1T}$, that captures the distribution of longitudinally polarized partons in transversely polarized hadrons.

Special interest in the wgt-function is caused by the observation that it gives rise to a $\cos(\varphi-\phi_S)$-modulation in the single-spin asymmetry (SSA) in the $p^\uparrow+p\to W/Z$ reaction. This asymmetry has the same general structure as the famous Sivers asymmetry \cite{Sivers:1989cc}, but is caused by the P-violation of weak-forces, instead of the T-oddness of the Sivers function \cite{Collins:2002kn}. Such SSA can be measured in the STAR experiment at RHIC \cite{STAR:2015vmv}. To make a theoretical prediction for this SSA one needs the wgt-function, which can be extracted using lower-energy semi-inclusive deep-inelastic scattering (SIDIS) data. Such an extraction has been recently done in ref.\cite{Bhattacharya:2021twu}. However, the results of ref.\cite{Bhattacharya:2021twu} cannot be applied to high-energy data (see a discussion about the problems of the approach in the case of the Sivers SSA in ref. \cite{Echevarria:2020hpy}). So far, the only realization of TMD factorization that successfully describes polarized and unpolarized SIDIS and Drell-Yan-type processes, at low and high-energy simultaneously, is the approach based on the $\zeta$-prescription \cite{Scimemi:2018xaf}. The $\zeta$-prescription is a particular solution of TMD evolution equations  which allows a consistent usage of TMD evolution without implementation (or with partial implementation) of collinear matching. In refs. \cite{Bury:2021sue, Bury:2020vhj} this approach was used for a N$^3$LO extraction of the Sivers function using also data from STAR.

The main aim of this work is to extract the wgt-function from the available SIDIS data and make a prediction for the $\sim\cos(\varphi-\phi_s)$-SSA in the kinematics of the STAR experiment. Additionally, we would like to test the recent NLO computation of the wgt-function  \cite{Rein:2022odl} (in a Wandzura-Wilczek-type approximation), and estimate the size of the twist-three contribution. The paper is structured as follows: In section \ref{sec:theory} we review the theory framework and define the model for the wgt-function. In section \ref{sec:comparision} we analyse the SIDIS data and determine the free parameters of our ansatz. Finally, in section \ref{sec:SSA}, we define the $\cos(\varphi-\phi_s)$-SSA induced by the weak bosons and make a prediction for the STAR experiment.

\section{Theoretical framework}
\label{sec:theory}

In this section, we summarize the TMD factorized expressions required for the present analysis. The general setup follows the one used in ref. \cite{Scimemi:2019cmh, Vladimirov:2019bfa, Bury:2020vhj, Bury:2021sue, Bury:2022czx}. The required input for polarization-independent elements is taken from the global extraction \cite{Scimemi:2019cmh}. A summary of the perturbative order used is given in table \ref{tab:PT-summary}. Our model for the wgt-function is discussed in subsection \ref{sec:wgt-model}.

\subsection{Transverse-longitudinal asymmetry in SIDIS}

Semi-inclusive deep-inelastic scattering is the reaction
\begin{eqnarray}
\ell(l)+H(P,S)\to \ell(l')+h(p_h)+X,
\end{eqnarray}
where $\ell$ is a lepton, $H$ and $h$ are target and produced hadron. The momentum of the intermediate photon is $q=l-l'$ ($Q^2=-q^2$). The kinematics of SIDIS is described by the standard variables
\begin{eqnarray}\nn
x&=&\frac{Q^2}{2 (qP)},\quad y=\frac{(qP)}{(lP)},\quad z=\frac{(p_hP)}{(qP)},\quad \gamma=\frac{2xM}{Q},
\\
\varepsilon&=&\frac{1-y-\frac{y^2\gamma^2}{4}}{1-y+\frac{y^2}{2}+\frac{y^2\gamma^2}{4}},
\end{eqnarray}
where $M^2=P^2$ is the mass of the target hadron. In addition to these longitudinal variables, one defines the transverse vector $p_{\perp}^\mu=g_\perp^{\mu\nu}p_{h,\nu}$ (and similar for $s_\perp$, $\ell_\perp$, etc), where
\begin{eqnarray}
g_\perp^{\mu\nu}&=&g^{\mu\nu}-\frac{1}{Q^2(1+\gamma^2)}\Big[
\\\nn &&2x^2 P^\mu P^\nu+2x(P^\mu q^\nu+q^\mu P^\nu)-\gamma^2 q^\mu q^\nu\Big],
\end{eqnarray}
is the projector to the plane orthogonal to $P^\mu$ and $q^\mu$.

The SIDIS cross-section is parameterized by many structure functions \cite{Bacchetta:2006tn}. The terms relevant in the present case are
\begin{eqnarray}
&&\frac{d\sigma}{dx dy d\psi dz dp_\perp^2}=\frac{\alpha_{\text{em}}^2}{xy Q^2}\frac{y^2}{2(1-\varepsilon)}\Big[F_{UU,T}
\\\nn &&\quad+|S_\perp|\lambda_\ell \sqrt{1-\varepsilon^2}\cos(\phi_h-\phi_S)F_{LT}^{\cos(\phi_h-\phi_S)}+...\Big],
\end{eqnarray}
where $\lambda_{\ell}$ is the helicity of the incoming lepton, and $\psi$ and $\phi$ are azimuthal angles \cite{Bacchetta:2004jz}. The arguments of these structure functions are $(x,z,Q^2,p_\perp^2)$. The transverse-longitudinal asymmetry is defined as
\begin{eqnarray}\label{def:ALT}
A_{LT}^{\cos(\phi_h-\phi_S)}=\frac{F_{LT}^{\cos(\phi_h-\phi_S)}}{F_{UU,T}}.
\end{eqnarray}

The theoretical description of SIDIS is performed in the hadronic frame, where the momenta of target and measured hadron are anti-parallel. In this frame the photon has a transverse momentum $q_T$ (orthogonal to $P^\mu$ and $p_h^\mu$), which is related to $p_\perp$ by
\begin{eqnarray}
q_T^2=\frac{p_\perp^2}{z^2}.
\end{eqnarray}
In this relation we have neglected corrections $\sim \gamma$ (see, f.i., ref.\cite{Scimemi:2019cmh}). In the limit $Q\to \infty$ and $q_T\ll Q$ the SIDIS cross-section obeys the TMD factorization theorem \cite{Bacchetta:2006tn, Collins:2011zzd}. According to it, the structure functions are
\begin{widetext}
\begin{eqnarray}
F_{UU,T}&=&|C_V(Q^2,\mu)|^2\sum_q e_q^2 
\int_0^\infty\frac{|b|d|b|}{2\pi}J_0\(\frac{|b||p_\perp|}{z}\)f_{1,q\ot H}(x,b;\mu,\zeta)D_{1,q\to h}(z,b;\mu,\bar \zeta)+\mathcal{O}\(\frac{q_T^2}{Q^2}\),
\\
F_{LT}^{\cos(\phi_h-\phi_S)}&=&|C_V(Q^2,\mu)|^2\sum_q e_q^2 
M
\int_0^\infty\frac{|b|^2d|b|}{2\pi}J_1\(\frac{|b||p_\perp|}{z}\)g_{1T,q\ot H}(x,b;\mu,\zeta)D_{1,q\to h}(z,b;\mu,\bar \zeta)+\mathcal{O}\(\frac{q_T^2}{Q^2}\),
\end{eqnarray}
\end{widetext}
where the $J$ are Bessel functions of the first kind and the $e_q$ are the electric quark charges. The function $f_1$ is the unpolarized TMD PDF. The function $D_1$ is the unpolarized TMD fragmentation function (FF). The coefficient function $|C_V|^2$ is perturbative, and cancels in the ratio (\ref{def:ALT}). We fix the factorization scales as $\mu=Q$ and $\zeta=\bar \zeta=Q^2$.

The unpolarized TMD distributions were defined and extracted in many works. The most recent studies describe hundreds of data points measured in Drell-Yan and SIDIS processes \cite{Bacchetta:2017gcc, Scimemi:2017etj, Scimemi:2019cmh, Bacchetta:2019sam, Bacchetta:2022awv}. Using these results one can predict the asymmetry (\ref{def:ALT}) up to the unknown wgt-function. In this work, we use the unpolarized input and evolution from \texttt{SV19} \cite{Scimemi:2019cmh} \footnote{
We use the so-called $m=0$-case of \texttt{SV19}. In this case, target-mass corrections are neglected. This is done in order to access low-Q data, where the target-mass corrections (in the leading power of the factorization theorem) push the kinematic variables into the unphysical region. The same choice was made in refs.\cite{Bury:2020vhj, Bury:2021sue}.
}.
\texttt{SV19} is based on the global fit of  data measured at Tevatron, LHC, FermiLab (for Drell-Yan), HERMES and COMPASS (for SIDIS). 
%The fit shows perfect agreement with data with $\chi^2/N_{pt} =1.1$ for Drell-Yan (with $N_{pt} = 457$) and $\chi^2/N_{pt} =0.95$ for SIDIS (with $N_{pt} = 582$). 
The \texttt{SV19} fit for unpolarized TMD distributions and the Collins-Soper kernel has been successfully used for subsequent extractions of other TMD distributions in refs. refs.\cite{Vladimirov:2019bfa, Bury:2020vhj, Bury:2021sue, Bury:2022czx, Zeng:2022lbo}.

The \texttt{SV19} analysis is made both at N$^2$LO and N$^3$LO. Here, we use the N$^3$LO version, which has full N$^3$LO TMD evolution. The unpolarized distributions are matched to collinear distributions at N$^2$LO. A summary of the perturbative input is given in table \ref{tab:PT-summary}.

\begin{table*}
%\begin{ruledtabular}
\begin{tabular}{|c||c|c|c|c||c|c|c|}
\hline
\multicolumn{2}{|c}{} & \multicolumn{3}{|c| |}{Evolution} & \multicolumn{3}{c|}{small-b matching}
\\\hline
Element & $C_V$ & $\Gamma_{\text{cusp}}$  & $\gamma_V$ & $\mathcal{D}_{\text{resum}}$ & $C \otimes f_1$ & $C \otimes d_1$ & $C \otimes g_{1}$ 
\\\hline
Order & $a_s^3$ & $a_s^4$ & $a_s^3$ & $a_s^3$ & $a_s^2$ & $a_s^2$ &  $a_s^1(\text{tw2-only})$ 
\\\hline
PDF/FF & \multicolumn{4}{c||}{} &
NNPDF3.1\cite{NNPDF:2017mvq} & DSS \cite{deFlorian:2014xna, deFlorian:2017lwf} & 
\begin{tabular}{@{}c@{}}NNPDF \cite{Nocera:2014gqa}\\ DSSV\cite{DeFlorian:2019xxt}\end{tabular}
\\\hline
\end{tabular}
\caption{\label{tab:PT-summary}
Summary of used orders of the perturbative series of different elements in the factorization formula. The perturbative order of the highest included term is shown. For the small-$b$ matching the order indicates the perturbative order of the coefficient function, and PDF/FF indicates the used collinear distribution.  This set up corresponds to N$^3$LL in the nomenclature of ref.\cite{Bacchetta:2019sam}. 
}
%\end{ruledtabular}
\end{table*}

\subsection{Evolution of TMD distributions}

The TMD distributions have two scaling parameters $\mu$ and $\zeta$. The corresponding evolution equations are \cite{Aybat:2011zv, Scimemi:2018xaf}
\begin{eqnarray}
\mu^2 \frac{d F(x,b;\mu,\zeta)}{d\mu^2}&=&\gamma_F(\mu,\zeta)F(x,b;\mu,\zeta), \label{eq:9}
\\
\zeta \frac{d F(x,b;\mu,\zeta)}{d\zeta}&=&-\mathcal{D}(b,\mu)F(x,b;\mu,\zeta), \label{eq:10}
\end{eqnarray}
where $F$ is any TMD distribution of leading twist, $\gamma_F$ is the TMD ultraviolet anomalous dimension and $\mathcal{D}$ is the Collins-Soper kernel. The anomalous dimension $\gamma_F$ is used at N$^3$LO ($a_s^4$ for cusp contribution, and $a_s^3$ for the rest) \cite{Gehrmann:2010ue}. The Collins-Soper kernel is generally nonperturbative \cite{Vladimirov:2020umg}, and is defined below. The solution of equation (\ref{eq:9}) and (\ref{eq:10}) is
\begin{eqnarray}\label{F=RF}
&&F(x,b;\mu_f,\zeta_f)
\\\nn &&=\exp\[\int_P\(\gamma_F(\mu,\zeta)\frac{d\mu}{\mu}-\mathcal{D}(b,\mu)\frac{d\zeta}{\zeta}\)\]
F(x,b;\mu_i,\zeta_i),
\end{eqnarray}
where $P$ is any path connecting the points $(\mu_f,\zeta_f)$ and $(\mu_i,\zeta_i)$. 

One of the central questions of TMD phenomenology is the choice for the initial scales $(\mu_i,\zeta_i)$ (we remind that $(\mu_f,\zeta_f)=(Q,Q^2)$). We fix these scales in accordance with the $\zeta$-prescription
\begin{eqnarray}\label{def:optimalScale}
(\mu_i,\zeta_i)=(\mu_0,\zeta_{\mu_0}(b)),
\end{eqnarray}
where $\mu_0$ can be any scale. The function $\zeta_{\mu}(b)$ is defined by the equation \cite{Vladimirov:2019bfa}
\begin{eqnarray}\label{def:zeta-line}
\gamma_F(\mu,\zeta_\mu(b))=2\mathcal{D}(b,\mu)\frac{d\ln \zeta_\mu(b)}{d\ln \mu^2},
\end{eqnarray}
with the boundary condition $\gamma_F(\mu_0,\zeta_{\mu_0}(b))=0$, where $\mu_0$ is defined implicitly by $\mathcal{D}(\mu_0,b)=0$. This boundary condition corresponds to the case that the line $(\mu,\zeta_\mu(b))$ passes through the saddle point of the evolution potential \cite{Scimemi:2018xaf}. Equation (\ref{def:zeta-line}) can be solved as expansion in $a_s$ for an arbitrary function $\mathcal{D}$ \cite{Vladimirov:2019bfa, Scimemi:2019cmh}. The expansion converges well, since $a_s(Q)\ll1$. We stress that this solution is valid for any values of $b$, since all non-perturbative physics is collected in the Collins-Soper kernel.

A TMD distribution at the scale (\ref{def:optimalScale}) is called optimal TMD distribution. By definition (\ref{def:zeta-line}), the optimal TMD distribution is independent on $\mu_0$. For that reason, it is usually denoted without scaling parameters
\begin{eqnarray}
F(x,b;\mu_0,\zeta_{\mu_0}(b))=F(x,b).
\end{eqnarray}
It is convenient to present equation (\ref{F=RF}) in the form
\begin{eqnarray}
F(x,b;Q,Q^2)=\(\frac{Q^2}{\zeta_{Q}(b)}\)^{-\mathcal{D}(b,Q)}F(x,b).
\end{eqnarray}
In this way, the TMD distribution is split into a factor dependent on $\mathcal{D}$ and the optimal TMD distribution that is independent of $\mathcal{D}$. This allows to theoretically decorrelate the (nonperturbative) evolution from the nonperturbative parton dynamics.

The Collins-Soper kernel parametrizes effects of the inter-quark interaction that is perturbative at small distances and nonperturbative at large distances. It is given by a vacuum  matrix element of a certain Wilson loop \cite{Vladimirov:2020umg}. The SV19 extraction defined the Collins-Soper kernel as follows
\begin{eqnarray}
\mathcal{D}(b,\mu)=\mathcal{D}_{\text{resum}}(b^*,\mu)+c_0 |b| b^*,
\end{eqnarray}
where $b^*=|b|(1+|b|^2/(2\text{GeV}^{-1})^2)^{-1/2}$, and $\mathcal{D}_{\text{resum}}$ is the resummed perturbative part of the Collins-Soper kernel \cite{Echevarria:2012pw}. The resummation is done at NNLO level. The necessary perturbative coefficients are taken from \cite{Vladimirov:2016dll,Li:2016ctv,Vladimirov:2017ksc}. The parameter $c_0$ is determined in \texttt{SV19} to be $c_0=0.0422\pm 0.011$.

It is important to emphasize that the $\zeta$-prescription completely decorrelates TMD distributions from evolution. Therefore, models for TMD distributions can be build without reference to the order in which evolution is treated. It is opposed to other schemes (such as Collins' scheme \cite{Aybat:2011zv, Collins:2011zzd}), where the order of evolution is tied to the order of the small-$b$ expansion. This decorrelation allows us to use the best known evolution setup (N$^3$LO) together with incomplete small-$b$ matching for the wgt-function (see next section).

\subsection{Model for the optimal worm-gear T-function}
\label{sec:wgt-model}

The wgt-function is a nonperturbative function of the two variables $(x,b)$ for each active flavor. Given the quality of the present data, it is practically impossible to extract the wgt-function without imposing strong model assumptions. To construct a realistic ansatz we use the small-$b$ expansion that relates the asymptotic behavior of the wgt-function to helicity distributions and twist-three distributions \cite{Kanazawa:2015ajw, Scimemi:2018mmi}.

\begin{figure}[t]
\begin{center}
\includegraphics[width=0.4\textwidth]{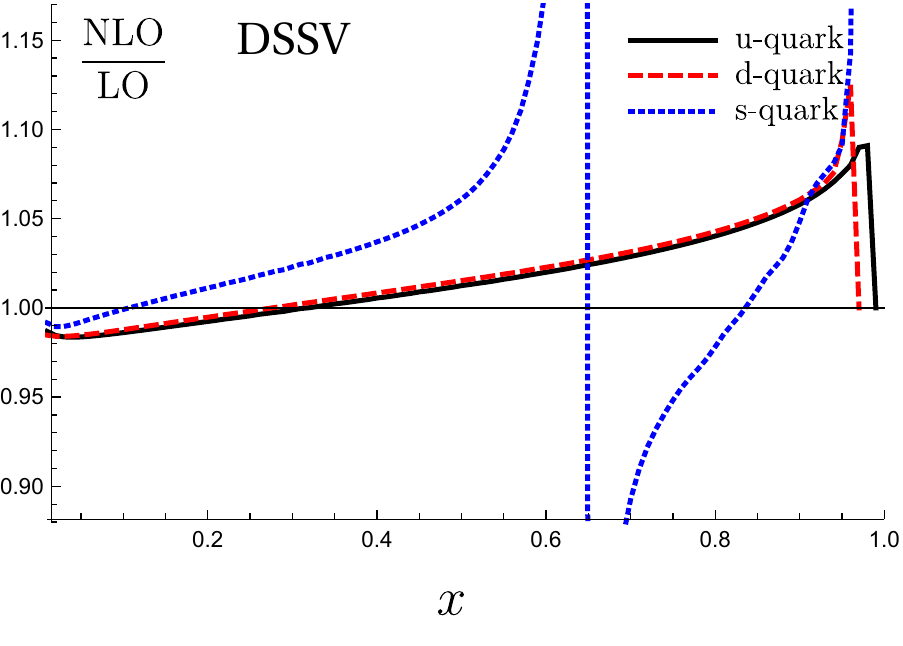}
\caption{\label{fig:NLO->LO} Comparison of NLO and LO twist-two approximations for the optimal wgt-function at $b=0.1$GeV$^{-1}$. 
%The left (right) plot shows the comparison made with NNPDF (DSSV) helicity distribution.
The comparison is shown for DSSV helicity distributions.}
\end{center}
\end{figure}

At small value of $b$, the optimal wgt-function has the form
\begin{widetext}
\begin{eqnarray}\label{sec:small-b}
g_{1T,q}^\perp(x,b)&=&\sum_{f}\, x\int_x^1 \frac{dy}{y} C^{\text{tw2}}_{q\ot f}\(\frac{x}{y},\mu_{\text{OPE}}\)g_{1f}(y,\mu_{\text{OPE}})+
\sum_{f}[C_{q\ot f}^{\text{tw3}}\otimes T_f](x)+\mathcal{O}(b^2),
\end{eqnarray}
where the first and second term represent the twist-two and the twist-three contributions, correspondingly. The summation index $f$ runs over all active flavors.

The twist-two contribution contains the helicity distribution $g_1$. The NLO coefficient functions (in the $\zeta$-prescription) read \cite{Rein:2022odl}
\begin{eqnarray}
C^{\text{tw2}}_{q\ot q}(x,\mu)&=&1+
a_s(\mu)C_F\[\mathbf{L}_b(2\ln x-4\ln(1-x)-1-2x)-2(1-x)-2\ln x-\frac{\pi^2}{6}\]+\mathcal{O}(a_s^2),
\\
C^{\text{tw2}}_{q\ot g}(x,\mu)&=&a_s\[-\mathbf{L}_b(\ln x+2-2x)+1-x+\frac{1}{2}\ln x\]+\mathcal{O}(a_s^2),
\end{eqnarray}
\end{widetext}
where $a_s=g^2(\mu)/(4\pi)^2=\alpha_s/(4\pi)$ and $\mathbf{L}_b=\ln(\mu^2 b^2 e^{2\gamma_E}/4)$. The usage of NLO is important because it gives access to the leading logarithm, and thus accounts for the LO effects of the evolution. The perturbative convergence of this series is good. In particular, the difference between LO and NLO is only of order $5$\%, as demonstrated in fig.\ref{fig:NLO->LO}.

The twist-three term contains twist-three quark-gluon-quark distributions $T$ and $\Delta T$ already at LO. At NLO, there is also the contribution of three-gluon distributions $G_\pm$ and $Y_\pm$, and a singlet quark mixing term. The explicit expression for $[C^{\text{tw3}}\otimes T]$ is rather long \cite{Rein:2022odl}, and it is not important in this context. At present, the twist-three functions are totally unknown, and thus this part of the small-$b$ expansion does not provide any useful information.

The expansion (\ref{sec:small-b}) serves as the base for our ansatz. As the first approximation, we neglect the twist-three contribution (Wandzura-Wilczek-type of approximation), and replace the large-b terms by the unknown function $f_{\text{NP}}$. The result reads
\begin{eqnarray}\label{ansatz}
&&g_{1T,q}^\perp(x,b)
\\\nn &&
=
\sum_{f}\, x\int_x^1 \frac{dy}{y} C^{\text{tw2}}_{q\ot f}\(\frac{x}{y},\mu_{\text{OPE}}\)g_{1f}(y,\mu_{\text{OPE}})\, f_{\text{NP}}(b).
\end{eqnarray}
In general, the function $f_{\text{NP}}$ depends also  on $x$ and flavor. In the present study, we neglect these dependences. The function $f_{\text{NP}}$ must vanish at $b\to\infty$ and its Taylor series at $b=0$ should contain only powers of $b^2$. This is a popular strategy for construction of a fitting ansatz, e.g., it is widely used for the extraction of unpolarized distributions.

The function $f_{\text{NP}}$ is to be determined from the data. We use the following two-parameter ansatz
\begin{eqnarray}
f_{\text{NP}}(b)=\frac{\lambda_2}{\cosh(\lambda_1 b)}.
\end{eqnarray}
This ansatz shows Gaussian behavior at $b\sim0$ and exponential decay at $b\to \infty$. Such a shape is supported by studies of the unpolarized distributions \cite{Scimemi:2017etj, Scimemi:2019cmh, Vladimirov:2019bfa}. At $\lambda_2=1$ the ansatz (\ref{ansatz}) reproduces the pure twist-two approximation. The deviation of the parameter $\lambda_2$ from unity simulates the eliminated twist-three part.

The collinear PDF $g_1$ is an essential input. In our model, it completely dictates the shape of the wgt-function. We have considered the two latest extractions of this function, performed by the NNPDF collaboration \cite{Nocera:2014gqa}, and by de Florian et al. \cite{DeFlorian:2019xxt}. For shortness we refer to these PDFs as NNPDF and DSSV, correspondingly. For the scale of the operator product expansion, we use the popular expression
\begin{eqnarray}
\mu_{\text{OPE}}=\frac{2e^{-\gamma_E}}{|b|}+2\text{GeV}.
\end{eqnarray}
With this choice the logarithms $\mathbf{L}_b$ vanish at $b\to0$ and $\mu_{\text{OPE}}\gg \Lambda_{\text{QCD}}$. So, the ansatz (\ref{ansatz}) is regular in the full range of $b$.

\section{Comparison with data}
\label{sec:comparision}

\subsection{Overview of experimental measurements}

There are three available measurements of $A_{LT}^{\cos(\phi_h-\phi_S)}$ asymmetry -- by HERMES \cite{HERMES:2020ifk}, COMPASS \cite{COMPASS:2016led,Parsamyan:2018evv}, and Jefferson Laboratory \cite{JeffersonLabHallA:2011vwy}.

The HERMES collaboration reported a large set of data for $A_{LT}^{\cos(\phi_h-\phi_S)}$ measured in $\pi^\pm$ and $K^\pm$ production at the proton target \cite{HERMES:2020ifk}. These data have a 4-dimensional binning in $(Q,x,z,p_\perp)$. Due to it the selection of the points that satisfy the TMD factorization is straightforward. 
%The reported systematic ??? uncertainty (due to the dilution and depolarization effects) is $7.8\%$. 

The COMPASS collaboration presented $A_{LT}^{\cos(\phi_h-\phi_S)}$ measured for a proton target with unidentified charged hadrons $h^\pm$ \cite{COMPASS:2016led,Parsamyan:2018evv}. These data have one-dimension binning (in $x$, $z$ or $p_\perp$), and split into three subsets with $z\in[0.1,0.2]$, $z\in[0.2,1]$ and $z\in[0.1,1]$.  For our analysis we have selected the $z\in[0.2,1]$ subset, because it contains most points in the TMD factorization region. We consider all three binning variants. This can lead to some double-counting of data, but in view of the large experimental uncertainties we made no effort to correct for it.
%The possible problem of double-counting of data points is compensated by their large uncertanties.
%The systematic ??? uncertainty is $3\%$.

Finally, eight points were reported by JLab Hall A \cite{JeffersonLabHallA:2011vwy}, for produced $\pi^\pm$ and a neutron target. These data points are measured at very low energy, and thus can be hardy described by any factorization approach. Therefore, we do not include them into the fit, but show the extrapolation of our results in fig.\ref{fig:JLab}.

To select the points which belong to the kinematic range of TMD factorization, we use the general strategy of \texttt{SV19}-fits. We introduce the variable $\delta=p_\perp/(zQ)$, where we set the kinematic variables to their mean values. The standard choice is to use the data with $\delta<0.25$ and $Q^2>4$GeV$^2$ \cite{Scimemi:2017etj, Scimemi:2019cmh, Bacchetta:2019sam,Vladimirov:2019bfa}. This choice securely guaranties the applicability of TMD factorization. However, in this case one finds only 23 points, many of which are the only representatives of a set of points and are located at the boundary of phase space. In other words, this strict selection criterion cannot be used for a meaningful analysis. Thus, we had to soften the demands and require
\begin{eqnarray}
\langle Q^2\rangle>2\text{GeV}^2,\qquad \delta<0.35.
\end{eqnarray}
With such a cut one has to expect the contribution from power corrections to be up to 30-50\% for points with the lowest $Q$ and the largest $\delta$. However, the statistical uncertainties of most of the data points are larger, which allows us to use these data points without extreme tension. Altogether we then have 70 points: 44 from the HERMES measurement (11 for each reaction), and 26 from the COMPASS measurement. The majority of data-points (49 out of 70) has $x\in[0.1,0.2]$, and only 5 points have $x>0.3$ (COMPASS).

\subsection{Comparison with data}

\begin{figure}[t]
\begin{center}
\includegraphics[width=0.45\textwidth]{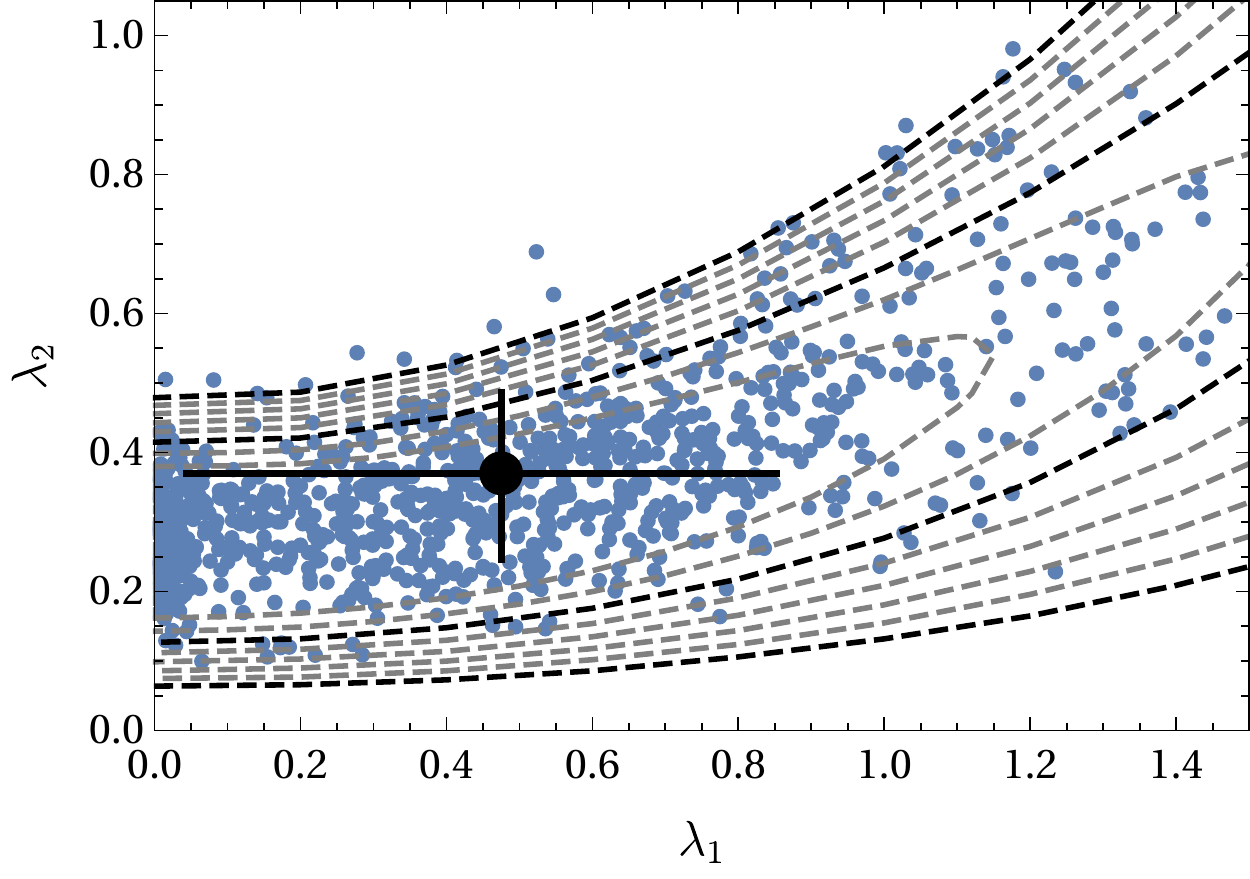}
\end{center}
\caption{\label{fig:lambda12} Distribution of values $\lambda_{1,2}$ in the bootstrapping ensemble. The black dot with bars indicate the mean value and its uncertainties. The line show the contours of equal $\chi^2/N_{\text{np}}$ (for central values), starting with $\chi^2/N_{\text{np}}=0.93$ till $\chi^2/N_{\text{np}}=1$.}
\end{figure}

The quality of agreement between the data and the model is determined with the $\chi^2$-test function. It is defined as
\begin{eqnarray}
\chi^2&=&\sum_{\substack{s\in \\ \text{data-sets}}} \sum_{\substack{i,j\in \\ \text{data-points}}} (t_i-a_i)V_{s,ij}^{-1}(t_j-a_j),
\end{eqnarray}
where $s$ runs over all independent data sets and $i,j$ over all points in $s$. Here, $t_i$ and $a_i$ are the theory prediction and the measured value for the point $i$. The covariance matrix $V_s$ is defined for each data set as
\begin{eqnarray}
V_{ij}=\delta_{ij}\sum_{l=1}^{N_{\text{uncor.}}}(\sigma_i^{(l),\text{uncor.}})^2+\sum_{l=1}^{N_{\text{cor.}}}\sigma_i^{(l),\text{cor.}}\sigma_j^{(l),\text{cor.}},
\end{eqnarray}
where $\sigma_i^{(l),\text{uncor.}}$ and $\sigma_i^{(l),\text{cor.}}$ are uncorrelated and correlated uncertainties of the $i$'th measurement. This definition takes into account the nature of experimental uncertainties, and gives a faithful estimate of the agreement between the experimental data and the theory prediction.

The evaluation of the theory prediction for a given set of model parameters is made by \texttt{artemide}. The code for the wgt-function and related processes has been added to \texttt{artemide} and is available at \cite{artemide}. Evaluation of the $\chi^2$ function and processing of data were performed using the Python interface for artemide, which is (together with all programs used for the current fit) available at \cite{dataProcessor}. The minimization of $\chi^2$ is made with the \texttt{iminuit} package (MINUIT2) \cite{iminuit}.

\begin{table}[b]
\begin{center}
\begin{tabular}{c||c|c|c|c||c|c|}
\cline{2-7}
&\multicolumn{4}{c||}{Hermes}
& \multicolumn{2}{c|}{Compass}
\\\cline{2-7}
& $\pi^+$ & $\pi^-$ & $K^+$ & $K^-$ & $h^+$ & $h^-$
\\\hline
\multicolumn{1}{|l||}{$N_{\text{pt}}$} & 11 & 11 & 11 & 11& 14 & 12
\\\hline
\multicolumn{1}{|l||}{$\chi^2/N_{\text{pt}}$} & 1.29  & 0.68  & 1.52 & 1.09 & 0.46 & 0.59  
\\\hline
\end{tabular}
\caption{\label{tab:chi2} The values of $\chi^2/N_{\text{pt}}$ for individual data sets. The values are presented for the DSSV case. For the NNPDF case the values differ by $\pm 0.02$.}
\end{center}
\end{table}

The uncertainty of the extraction is estimated using the bootstrap method. For that we generated an ensemble of pseudo-data replicas (with 1000 replicas) distributed according to the measured uncertainties \cite{Ball:2008by}. For each replica in the ensemble we performed the minimization procedure and obtained an ensemble of parameters $\lambda_{1,2}$. The ensemble has a very asymmetric form, see fig.\ref{fig:lambda12}. Therefore, we compute the asymmetric 68$\%$C.I. and present it as the uncertainty.

The resulting values for the NNPDF helicity PDF are
\begin{eqnarray}
\text{NNPDF}:\quad \lambda_1=0.52_{-0.42}^{+0.37},\qquad \lambda_2=0.41_{-0.12}^{+0.14}~,
\end{eqnarray}
For DSSV helicity PDF we have
\begin{eqnarray}
\text{DSSV}:\quad \lambda_1=0.47_{-0.43}^{+0.38},\qquad \lambda_2=0.37_{-0.12}^{+0.12}~.
\end{eqnarray}
The values of $\chi^2$ at the mean values is $\chi^2/N_{\text{pt}}=0.92$ (with $N_{\text{pt}}=70$) for both PDF cases. $\chi^2/N_{pt}$ for the different reaction channels is given in table \ref{tab:chi2}.

\begin{figure}[t]
\begin{center}
\includegraphics[width=0.49\textwidth]{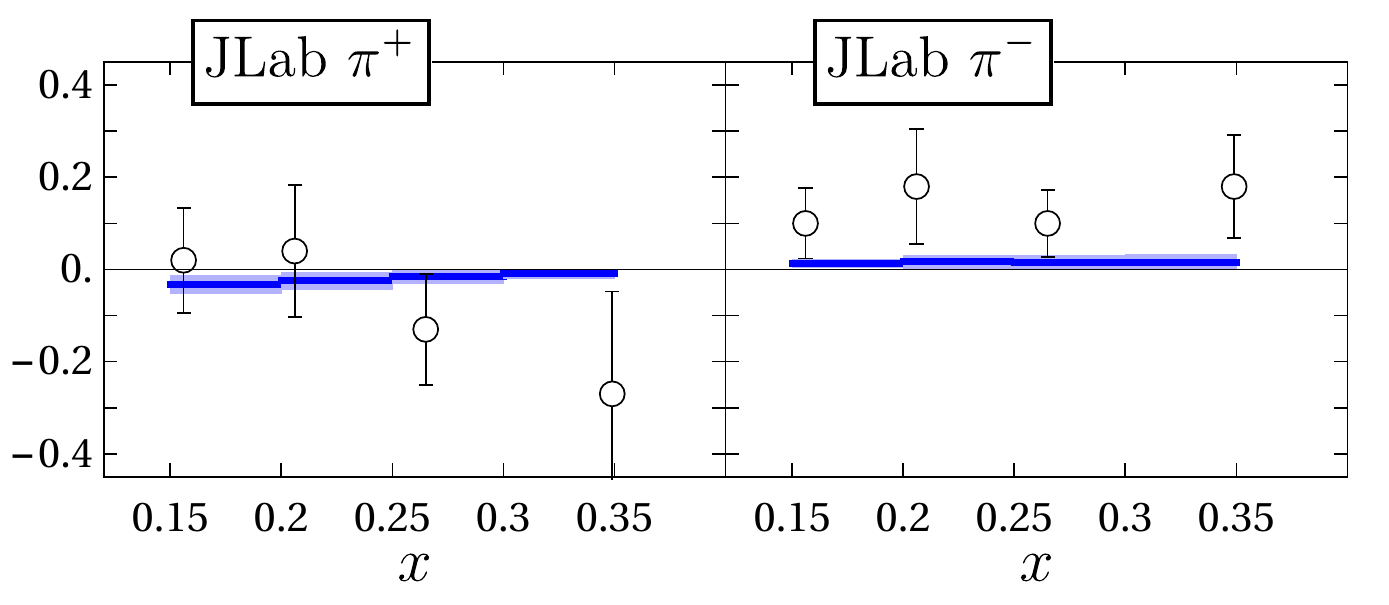}
\caption{\label{fig:JLab} The JLab data for $A_{LT}^{\cos(\phi_h-\phi_S)}$. The lines and shaded areas show the theory predictions.}
\end{center}
\end{figure}

Both helicity PDFs (NNPDF and DSSV) result in fits of comparable quality and give similar parameters for the wgt-functions. The distribution of $\chi^2$'s for the particular measurements is almost the same (differing only in the third digit). Clearly, this is due to the poor quality of the data, which cannot distinguish the $5-10$\% difference in the input PDF. Therefore, we continue with DSSV PDFs, assuming that NNPDF gives analogous results.

We have also performed a fit including the uncertainty of PDFs. For that we fitted $\lambda_{1,2}$ with randomly selected replicas of PDFs. The procedure is described in ref.\cite{Bury:2022czx}. The result is in very good agreement with the central PDF fit, increasing the final uncertainty band by $3$\% only. This is due to the fact that most data are localized in a small range of $x$. Therefore, the variation of PDF values is almost equivalent to the variation of normalization, which, in turn, is compensated by the variation of the parameter $\lambda_2$. 

In fact, the present data cannot truly constrain the wgt-function. In fig.\ref{fig:lambda12}, we show lines of equal $\chi^2/N_{\text{pt}}$ with the outmost line corresponding to $\chi^2/N_{\text{pt}}=1$. The area surrounded by these line extends up to $\lambda_1\sim 100$ (however, in this region, parameter $\lambda_1$ strongly correlates with $\lambda_2$). Such huge values correspond to a definitely unphysical wgt-function. Also, the  present data cannot safely exclude vanishing wgt-function. In particular, for  $\lambda_2=0$ we obtain $\chi^2/N_{\text{np}}=1.06$.

The value $\lambda_2=1$ corresponds to the pure twist-two approximation (Wandzura-Wilczek-type approximation). For this case, we get $\lambda_1=1.90^{+0.34}_{-0.43}$ with $\chi^2/N_{\text{pt}}=0.95$. Such a small difference between $\chi^2$'s with and without correction for twist-three does not allow us to estimate its size. However, if future better data confirmed the position of the $\chi^2$ minimum, the twist-three correction would be $\sim 50\%$ of the twist-two part.

\begin{figure*}
\begin{center}
\includegraphics[width=0.8\textwidth]{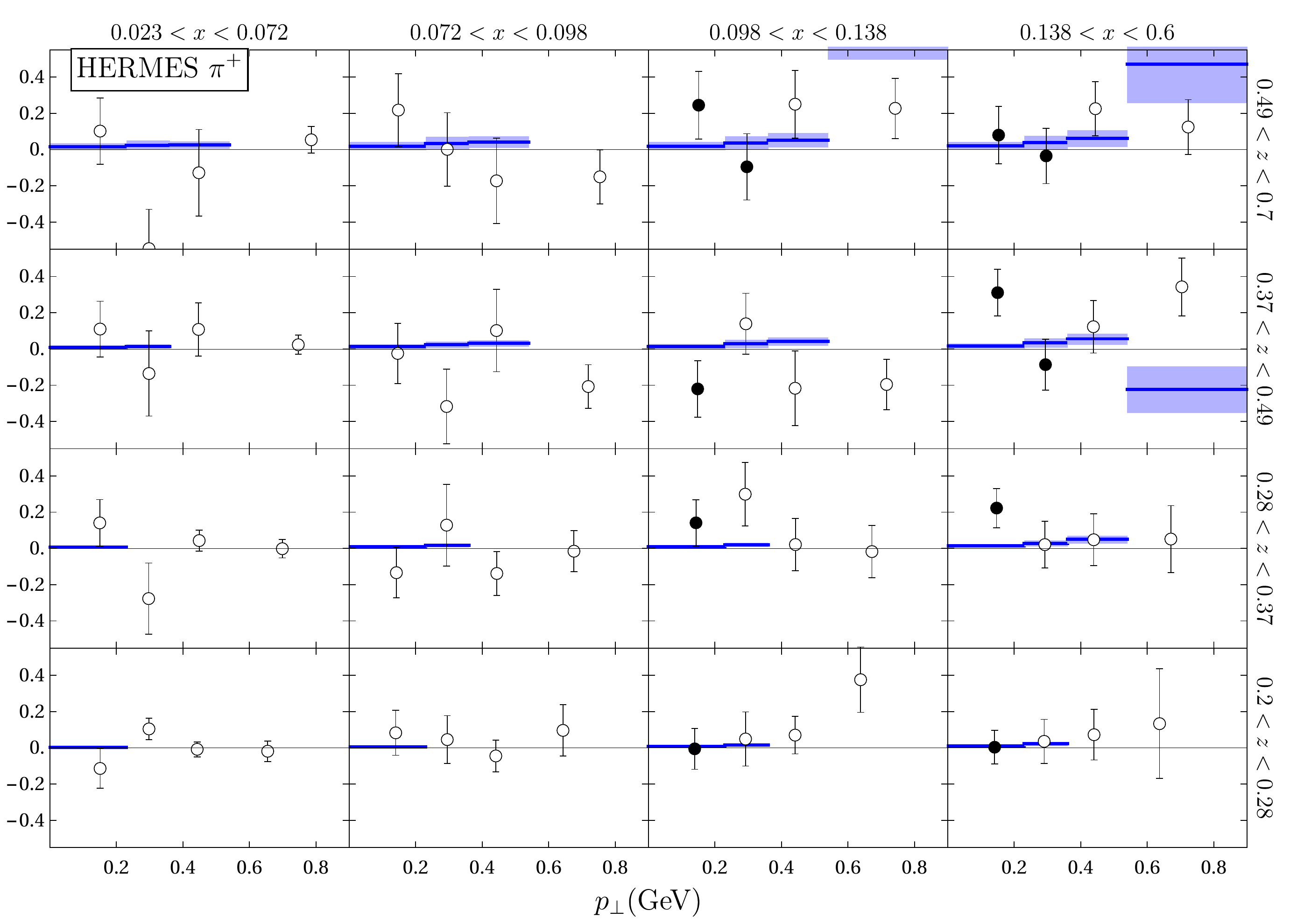}
\includegraphics[width=0.8\textwidth]{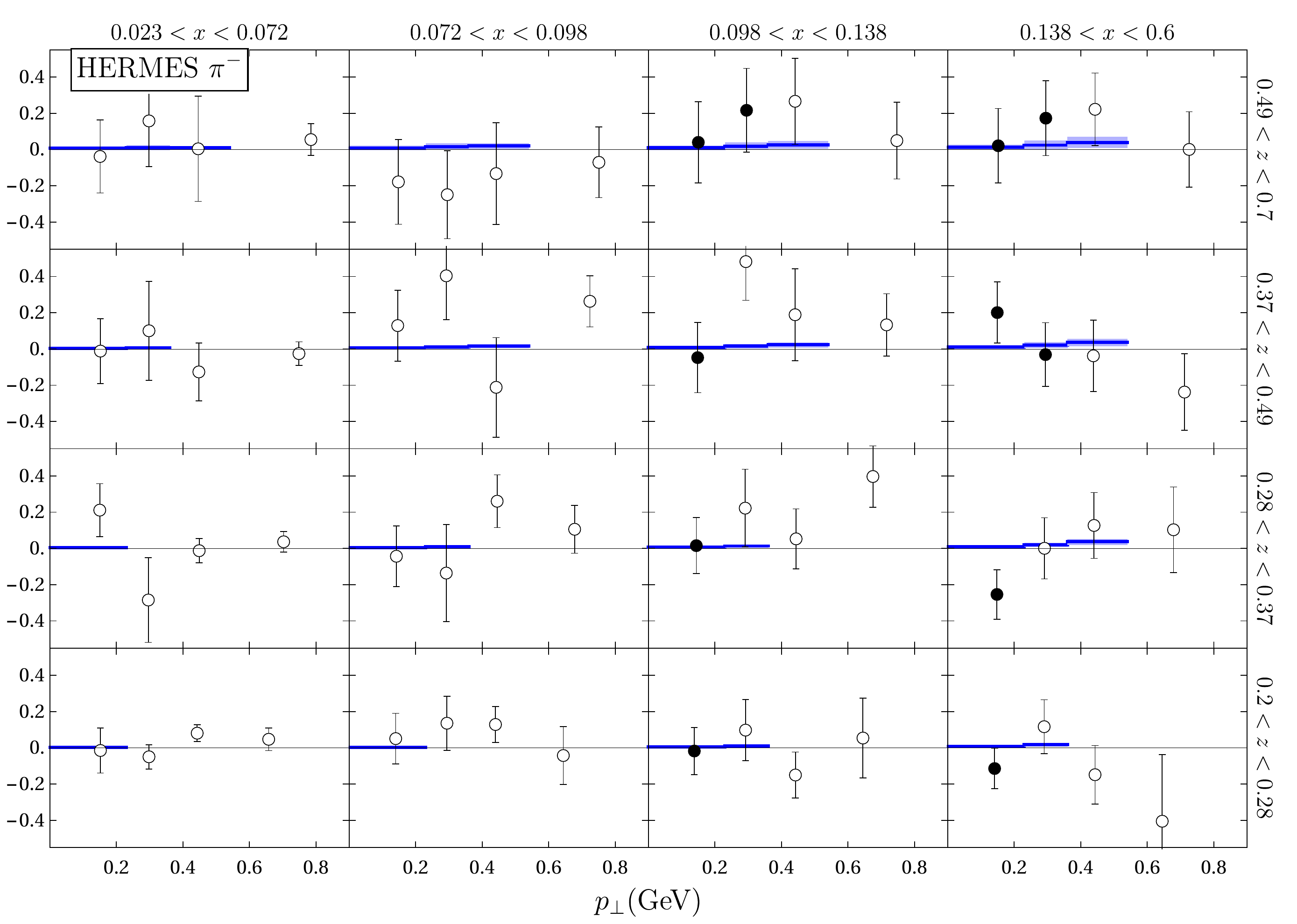}
\end{center}
\caption{\label{fig:HERMES:pi} The Hermes $\pi^+$ and $\pi^-$ data for $A_{LT}^{\cos(\phi_h-\phi_S)}$ for bins in $x$ and $z$. The lines and shaded areas show the theory prediction and its uncertainty for points with $\delta<0.75$. The filled points were included into the fit.}
\end{figure*}

\begin{figure*}
\begin{center}
\includegraphics[width=0.8\textwidth]{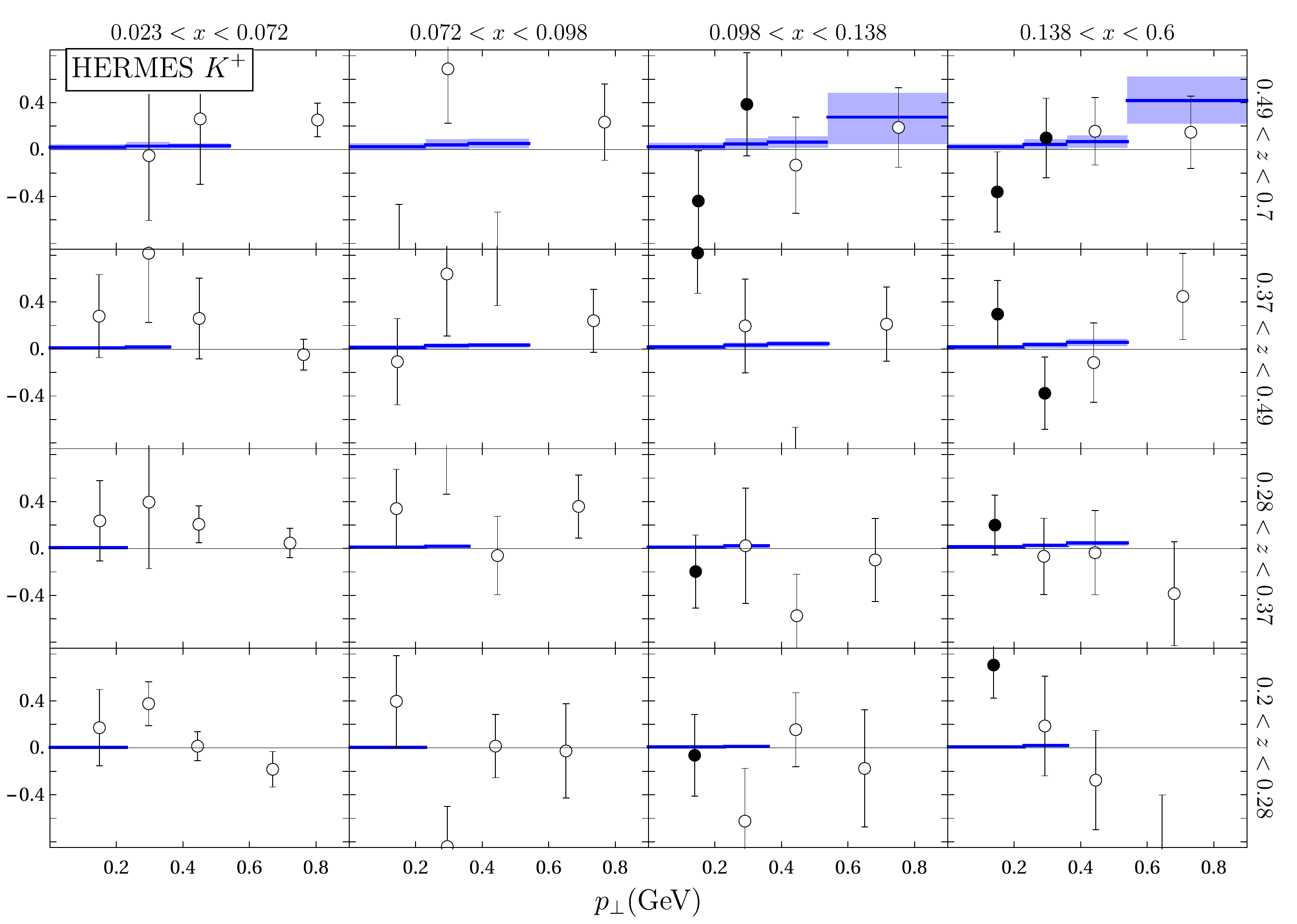}
\includegraphics[width=0.8\textwidth]{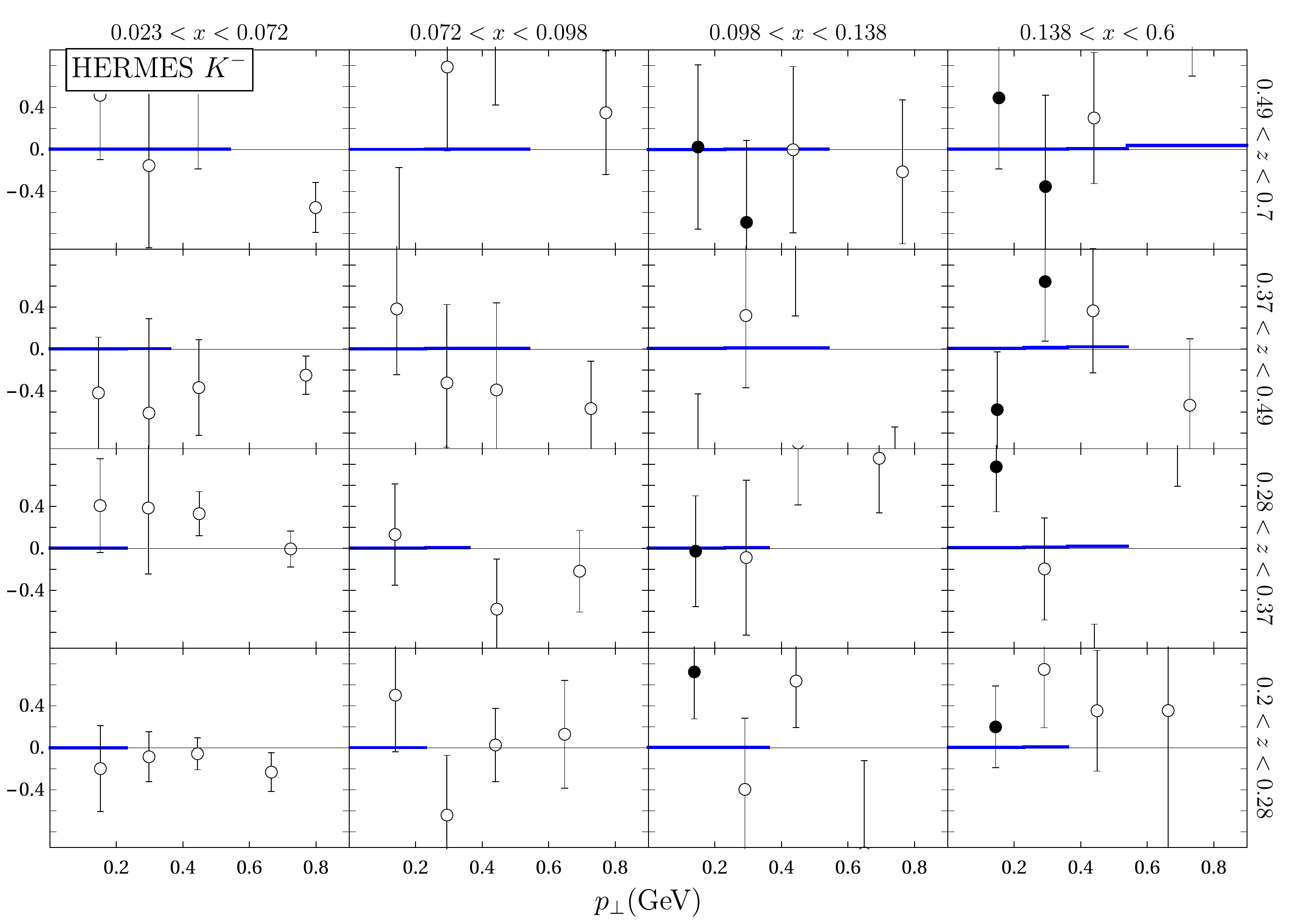}
\end{center}
\caption{\label{fig:HERMES:K} The Hermes $K^+$ and $K^-$ data for $A_{LT}^{\cos(\phi_h-\phi_S)}$ for bins in $x$ and $z$. The lines and shaded areas show the theory prediction and its uncertainty for points with $\delta<0.75$. The filled points were included into the fit.}
\end{figure*}

\begin{figure*}
\begin{center}
\includegraphics[width=0.6\textwidth]{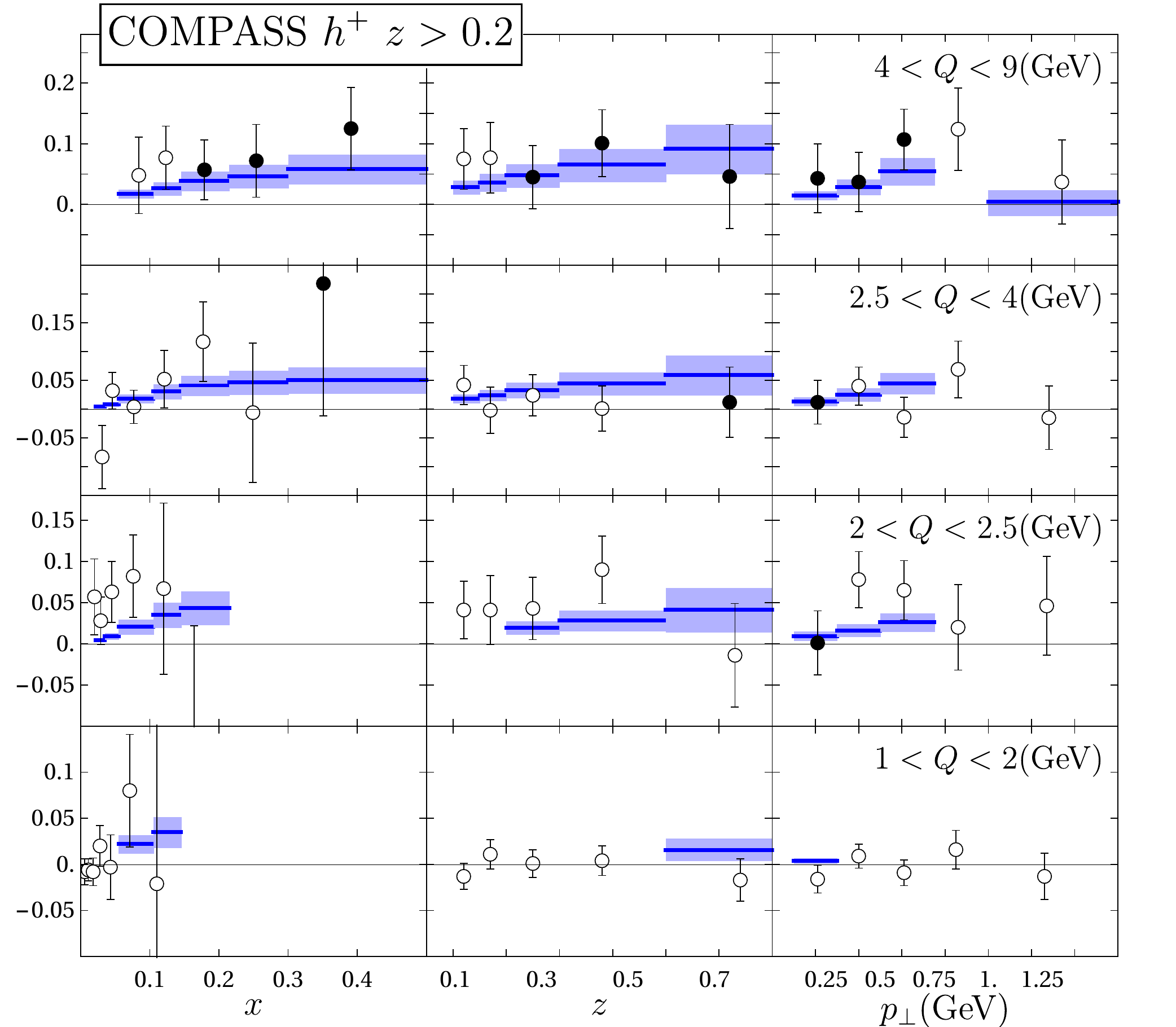}
\includegraphics[width=0.6\textwidth]{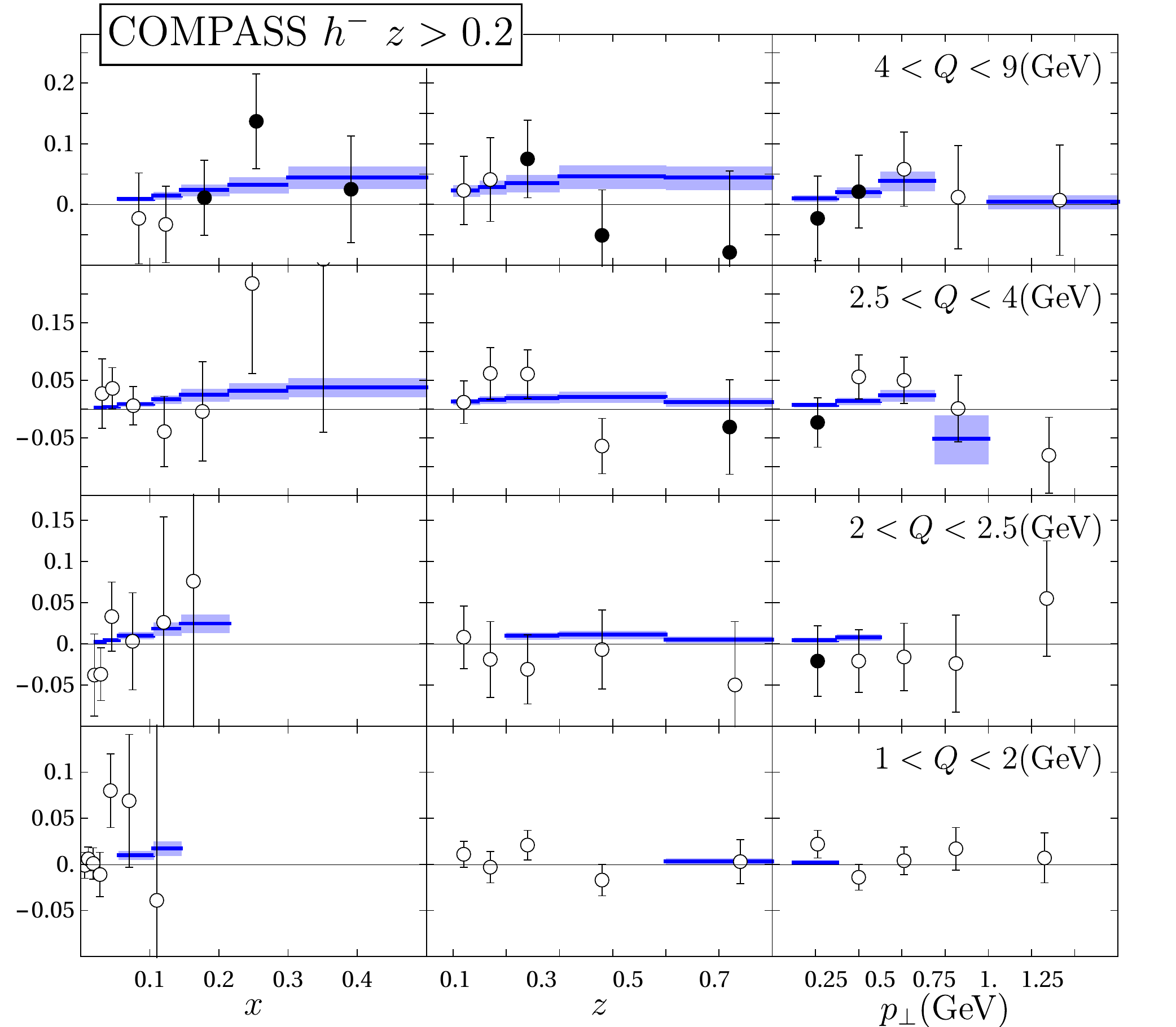}
\end{center}
\caption{\label{fig:COMPASS} The COMPASS $h^\pm$ data for $A_{LT}^{\cos(\phi_h-\phi_S)}$ in different bins. The lines and shaded areas show the theory prediction and its uncertainty for points with $\delta<0.75$. The filled points were included into the fit.}
\end{figure*}

A comparison of the theory prediction with the experimental data for all used data-sets is shown in fig.\ref{fig:HERMES:pi}, \ref{fig:HERMES:K}, and \ref{fig:COMPASS}. It is evident that the theory uncertainty is much smaller than the experimental uncertainty. The reason is the proportionality of $A_{LT}^{\cos(\phi_h-\phi_S)}$ to $p_\perp$. Thus, the asymmetry has very small absolute values for $p_\perp\sim 0.-0.4$GeV, where most data is localized. In fig.\ref{fig:JLab}, we show the comparison with JLab measurements \cite{JeffersonLabHallA:2011vwy}, which are done for $Q\sim 1.2-1.6$GeV, and were not included into the fit. The predicted asymmetry is generally much smaller than the uncertainty of the measurement.

\subsection{The fitted worm-gear T-function}

\begin{figure}[tb]
\begin{center}
\includegraphics[width=0.4\textwidth]{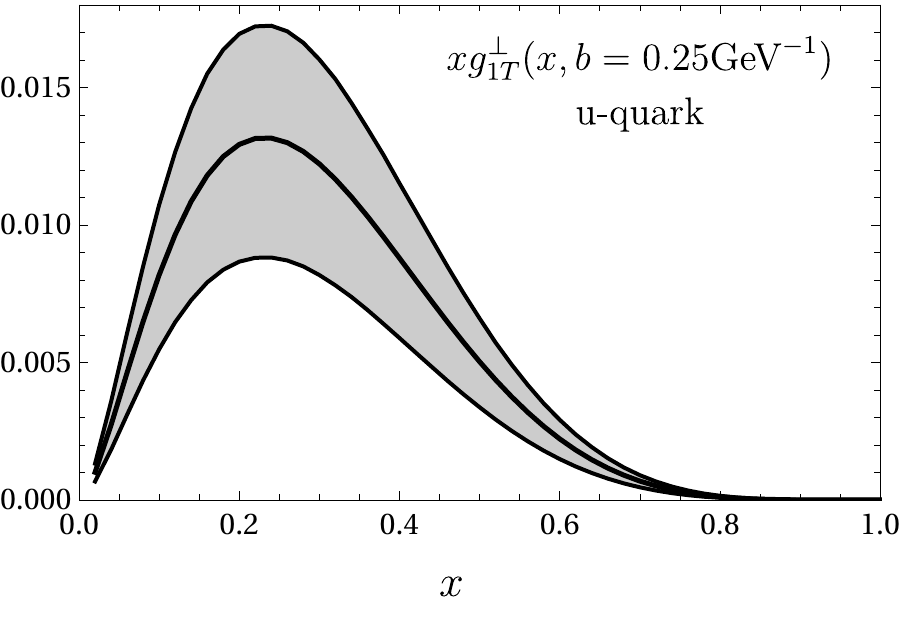}
\includegraphics[width=0.4\textwidth]{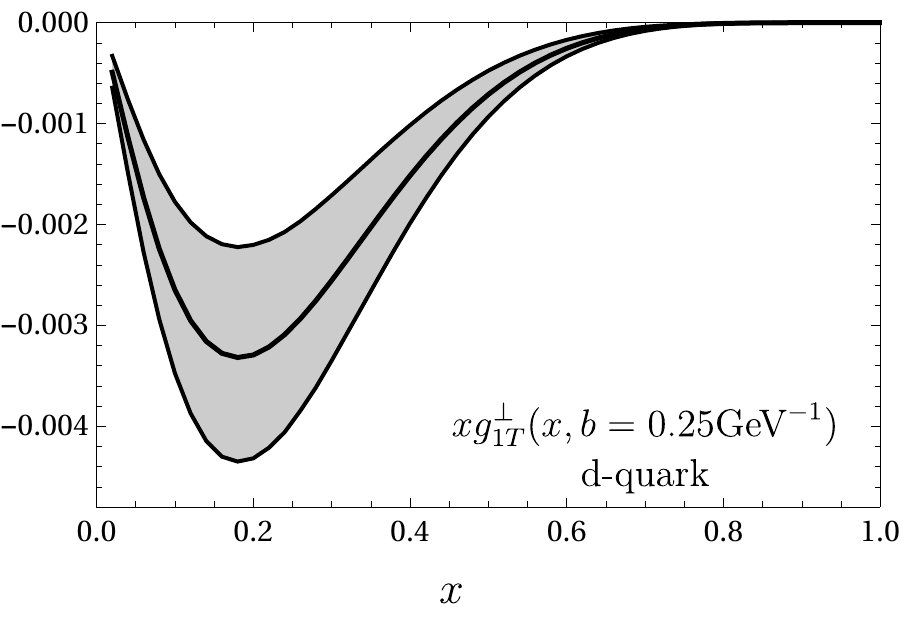}
\caption{\label{fig:wgt} The optimal wgt-function at $b=0.25$GeV$^{-1}$ for u and d-quarks. The plot is shown for DSSV helicity PDF. The NNPDF plots are very similar.}
\end{center}
\end{figure}

The extracted wgt-functions are shown in fig.\ref{fig:wgt} at $b=0.25$GeV$^{-1}$. Their shape agrees with the one extracted in ref.\cite{Bhattacharya:2021twu}, but the size is smaller by about a factor 4. A direct comparison with ref.\cite{Bhattacharya:2021twu} is not possible because the extraction in ref.\cite{Bhattacharya:2021twu} is made in a different scheme and in momentum space. 

The relative sign of the wgt-functions for $u$- and $d$- quark agrees with the large-$N_c$ estimate $g_{1T,u}^\perp \simeq -g_{1T,d}^\perp$ \cite{Pobylitsa:2003ty}. It is a consequence of the signs of the helicity PDFs. To test the sensitivity to the flavor-decomposition further, we introduced extra parameters in our ansatz. Namely, we fit $\lambda_2$ separately for $u$, $d$ and the rest of the flavors (sea). We obtain
\begin{eqnarray}
\lambda_1&=&0.33^{+0.35}_{-0.32},\quad
\lambda^u_2=0.40^{+0.13}_{-0.15},
\\\nn 
\lambda^d_2&=&3.1^{+2.5}_{-2.3},\quad
\lambda^{sea}_2=6.3^{+3.4}_{-4.1},
\end{eqnarray}
with $\chi^2/N_{\text{pt}}=0.89$. Clearly, the data constraints mainly the $u$-quark contribution. The other quarks (including the $d$-quark) remain largely unconstrained. The same conclusion has been reached in ref.\cite{Bhattacharya:2021twu}.

\section{Prediction for the weak-boson induced SSA}
\label{sec:SSA}

The Drell-Yan process is the reaction 
\begin{eqnarray}
h_1(p_1)+h_2(p_2)\to G(q)+X,
\end{eqnarray}
where $h$'s are colliding hadrons, and $G$ is the intermediate electro-weak gauge boson (detected by decay products). The hadronic tensor for this process reads
\begin{eqnarray}\label{DY:hadronT}
&&W_{\mu\nu}=
\\\nn 
&&\int\frac{d^4x}{(2\pi)^4} e^{-i(x\cdot q)}\sum_X \langle h_1,h_2|J_\mu^{\dagger}(x)|X\rangle\langle X|J_\nu(0)|h_1,h_2\rangle,
\end{eqnarray}
where
\begin{eqnarray}\label{weak-current}
&&J^\mu(x)=
\\\nn &&
\sum_{f,f'}\bar q_f(x)[g_R^G\gamma^\mu(1+\gamma^5)+g_L^G\gamma^\mu(1-\gamma^5)]_{ff'}q_{f'}(x),
\end{eqnarray}
with $\bar q$ and $q$ being quark fields, $g_{R(L)}^G$ being the coupling constants for right(left)-handed gauge bosons $G$, and indices $f$, $f'$ indicating the flavor. 

The full structure of the hadronic tensor (\ref{DY:hadronT}) including polarization and the resulting cross-sections is complicated. For the electro-magnetic current $G=\gamma$, it has been derived in ref.\cite{Arnold:2008kf}. For the  general current the complete structure has not yet been derived to the best of our knowledge. So far it was not needed since there was no possibility to measure the polarized DY process at energies compatible with the weak boson masses. However, the STAR experiment at RHIC published recently measurements of the Single-Spin Asymmetry (SSA) in Z/W-boson production \cite{STAR:2015vmv}, and plans to update this measurement in the nearest future. Also, there is a potential possibility to measure such processes at the fixed-target upgrade of LHCb \cite{Barschel:2015mka, Aidala:2019pit}. Therefore, it is interesting to look into the details of the polarisation dependence of Z/W-boson production.

As a prerequisite to this work, we have studied SSAs induced by weak currents (\ref{weak-current}) in the TMD factorization framework. Among many terms contributing to the cross-section we found the following
\begin{eqnarray}
&& \frac{d\sigma}{dQ^2 dy d\varphi dq_T^2}=
\frac{\alpha_{\text{em}}^2(Q)}{9sQ^2}\Big\{F_{UU}^1
\\\nn && +|s_T|\sin(\varphi-\phi_S)F_{TU}^1+|s_T|\cos(\varphi-\phi_S)\Delta F_{TU}^{1}+...\Big\},
\end{eqnarray}
where $F_{UU}^1$ is the unpolarized structure function, $F_{TU}^1$ is the Sivers structure function, and $\Delta F_{TU}^{1}$ is the structure function induced by parity-violating terms in the weak current. Here, $Q^2=q^2$, and $y$ and $q_T$ are the invariant mass and rapidity of the gauge boson. The variables $\varphi$ and $q_T$ are the angle and the absolute value of the transverse component of the gauge boson momentum measured in the center-of-mass frame. The spin $S$ is the spin of $h_1$. The ellipsis denotes contributions containing other DY structure functions.

TMD factorization yields the following expressions for SSA structure functions
\begin{widetext}
\begin{eqnarray}\label{def:FUT}
F_{TU}^1&=&-M|C_V(-Q^2,\mu)|^2\sum_{f,\text{ch.}} z_{ll'}^{\text{ch.}}z_{ff'}^{\text{ch.}}\Delta^{\text{ch.}}
\int_0^\infty \frac{b^2 db}{2\pi}J_1(b|q_T|)
\[f_{1T;f\ot h_1}^\perp \, f_{1;\bar f'\ot h_2}+f_{1T;\bar f\ot h_1}^\perp\, f_{1;f'\ot h_2}\],
\\\label{def:GUT}
\Delta F_{TU}^1&=&M|C_V(-Q^2,\mu)|^2\sum_{q,\text{ch.}} \bar z_{ll'}^{\text{ch.}}z_{ff'}^{\text{ch.}}\Delta^{\text{ch.}}
\int_0^\infty \frac{b^2 db}{2\pi}J_1(b|q_T|)
\[g_{1T;f\ot h_1} \, f_{1;\bar f'\ot h_2}-g_{1T;\bar f\ot h_1}\, f_{1;f'\ot h_2}\],
\end{eqnarray}
\end{widetext}
where $l$ and $l'$ indicate the measured leptons, and the arguments of the TMD distributions are $(x_1,b;\mu,\zeta)$ and $(x_2,b;\mu,\bar \zeta)$. The notation $\bar f$ indicates an anti-quark distribution with flavor $f$. The factors $z$ and $\Delta$ depend on the reaction channel. The possible channels are $\{\gamma\gamma, \gamma Z, ZZ, WW\}$, where $\gamma Z$ is the interference between $\gamma$- and $Z$- boson production amplitudes. The factors $z$ are defined as
\begin{eqnarray}
z^{GG'}_{ff'}&=&2(g_{R}^G g_{R}^{G'}+g_{L}^G g_{L}^{G'})_{ff'},
\\
\bar z^{GG'}_{ff'}&=&2(g_{R}^G g_{R}^{G'}-g_{L}^G g_{L}^{G'})_{ff'}.
\end{eqnarray}
The factors $\Delta$ are combinations of propagators of gauge bosons. The explicit expressions for $z_{ff'}$ and $\Delta_{ff'}$ can be found, e.g., in equations (2.26) and (2.27) of ref.\cite{Bury:2021sue}. The explicit expressions for the factor $\bar z$ are
\begin{eqnarray}
\bar z^{\gamma\gamma}_{ff'}&=&0,
\qquad \bar z^{\gamma Z}_{ff'}=-\delta_{ff'}\frac{|e_f|}{4s_Wc_W},
\\\nn
\bar z^{ZZ}_{ff'}&=&\delta_{ff'}\frac{4|e_f|s_W^2-1}{8s_W^2c_W^2},
\qquad \bar z^{WW}_{ff'}=-\frac{|V_{ff'}|^2}{4s_W^2},
\end{eqnarray}
where $e_f$ is the charge of field $f$, $s_W$ and $c_W$ are sine and cosine of the Weinberg angle, and $V_{ff'}$ is the relevant element of the Cabibbo-Kobayashi-Maskawa matrix.

Despite visual similarities between (\ref{def:FUT}) and (\ref{def:GUT}) the mechanisms generating the SSAs in $\Delta F_{TU}^1$ and $F_{TU}^1$ are totally different. The Sivers SSA is non-vanishing due to the T-oddness of the Sivers function, whereas $\Delta F_{TU}^1$ is non-vanishing due to the parity violation of weak interactions. That leads to the ``unusual'' minus sign between quark and anti-quark contributions in (\ref{def:GUT}).

\begin{figure}[t]
\begin{center}
\includegraphics[width=0.49\textwidth]{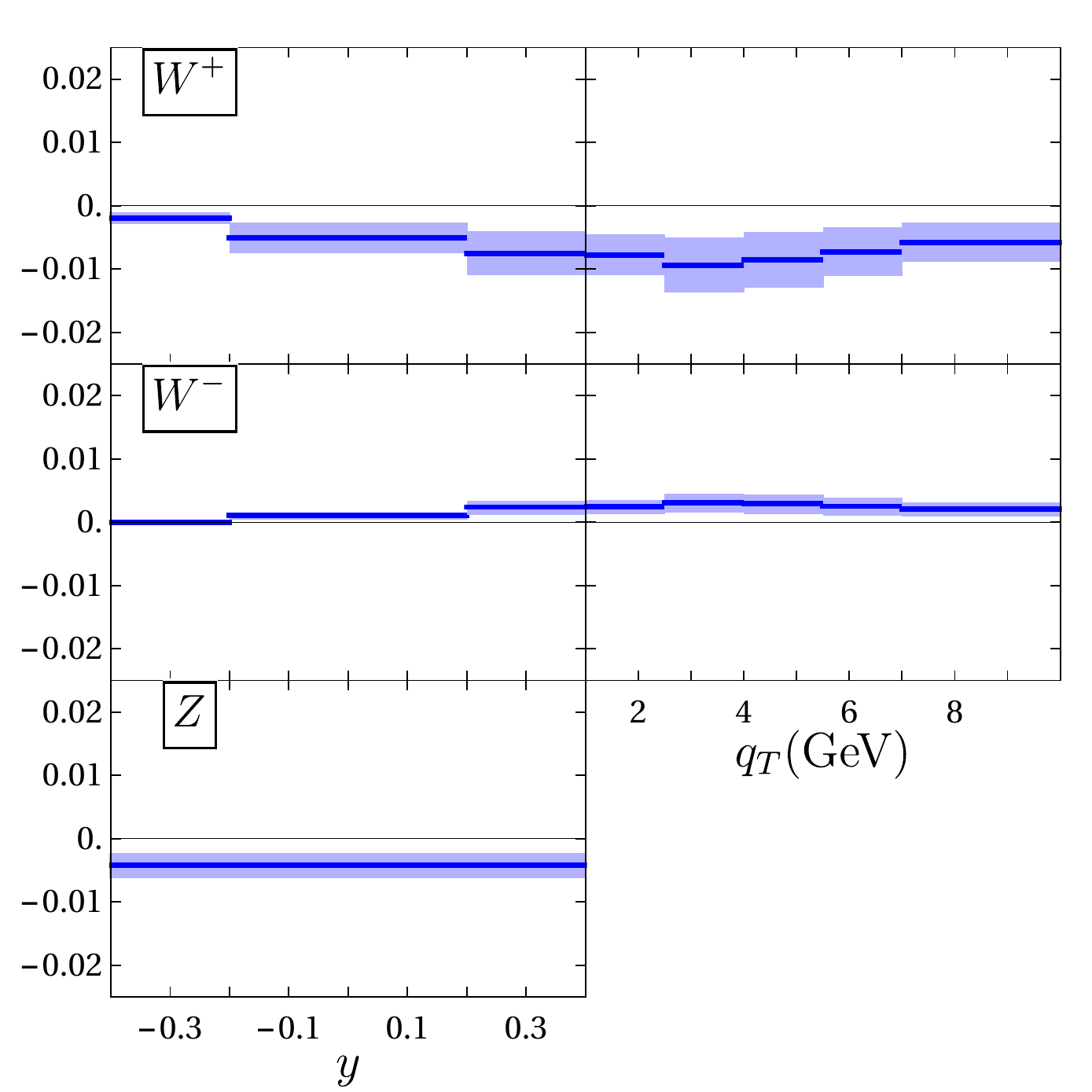}
\caption{\label{fig:STAR} The prediction for $\Delta F_{UT}^1/F_{UU}$ made in the STAR kinematics \cite{STAR:2015vmv}.}
\end{center}
\end{figure}

The structure function $F_{TU}^1$ is usually measured comparing the number of events in the left and right hemispheres (with respect to the polarization vector). The structure function $\Delta F_{TU}^1$ can be measured analogously comparing the number of events in the upper and lower hemispheres. Note that these contributions to the cross-section are orthogonal and thus do not contaminate each other. In fig.\ref{fig:STAR} we show the prediction for the asymmetry $\Delta F_{TU}^1/F_{UU}$ for the kinematics of the STAR experiment at RHIC, using the same binning as the Sivers measurement \cite{STAR:2015vmv}. While the predicted asymmetry is only of the order 1\%, our knowledge of the wgt-function is so poor that the effect could be larger by factor 2-3 without introducing any discrepancy with the present data.

\section{Conclusion}

We extracted the worm-gear-T (wgt-) function from the SIDIS data measured at COMPASS and HERMES. The analysis is done with N$^3$LO evolution and NLO matching (twist-two part only). The unpolarized elements of the factorization formula are taken from the global extraction made in ref.\cite{Scimemi:2019cmh} (made at N$^3$LL accuracy). The consistency of perturbative orders is guarantied by the $\zeta$-prescription. The results of the extraction as well as the code for the data analysis are available as a part of the \texttt{artemide}-library \cite{artemide}.

Our analysis clearly demonstrates that the present data cannot significantly restrict the wgt-function. There are two reasons for this. The first reason is that the modern data is localised in a rather narrow kinematic region. The second reason is that the theoretical cross-section is proportional to $p_\perp$, and thus has reduced sensitivity to the variations of wgt-function. In general, our results are in agreement with those of ref.\cite{Bhattacharya:2021twu}.

In addition, we made a prediction for the new SSA in the weak-boson production in Drell-Yan reactions, which could be measured at STAR at RHIC. This new SSA is similar to the Sivers SSA but with a $\cos(\varphi-\phi_S)$-modulation. On the theory side it is proportional to convolutions of the unpolarized and wgt TMD distributions. The predicted values are small, at the level of a few percents. 

\begin{acknowledgments}
We thank Rodolfo Sassot for DSSV grid tables. A.V. is funded by the \textit{Atracci\'on de Talento Investigador} program of the Comunidad de Madrid (Spain) No. 2020-T1/TIC-20204. A.V. is also supported by the Spanish Ministry grant PID2019-106080GB-C21. A.S. thanks the University of the Basque Country, Bilbao for hospitality. This work was partially supported by DFG FOR 2926 ``Next Generation pQCD for  Hadron  Structure:  Preparing  for  the  EIC'',  project number 430824754.
\end{acknowledgments}

\bibliography{bibFILE}
\end{document}